\documentclass[pra,twocolumn,amsmath,amssymb,floatfix]{revtex4-1}

\usepackage{graphicx}
\usepackage{dcolumn}
\usepackage{bm}
\usepackage{amsmath}
\usepackage{amssymb}
\usepackage{float}
\begin{document}

\title{Accurate prediction of clock transitions in a highly charged ion with complex electronic structure}
\author{C. Cheung$^1$, M. S. Safronova$^{1,2}$, S. G. Porsev$^{1,3}$, M. G. Kozlov$^{3,4}$, I. I. Tupitsyn$^{5,6}$, and A. I. Bondarev$^{6,5,3}$}
\affiliation{
$^1$Department of Physics and Astronomy, University of Delaware, DE, USA, \\
$^2$Joint Quantum Institute, NIST and the University of Maryland, College Park, MD, USA,\\
$^3$Petersburg Nuclear Physics Institute of NRC ``Kurchatov Institute'', Gatchina 188300, Russia,\\
$^4$St.~Petersburg Electrotechnical University ``LETI'', Prof. Popov Str. 5, 197376 St.~Petersburg, Russia,\\
$^5$Department of Physics, St. Petersburg State University, Ulianovskaya 1, Petrodvorets, St. Petersburg, 198504, Russia,\\
$^6$Center for Advanced Studies, Peter the Great St. Petersburg Polytechnic
    University, Polytekhnicheskaja 29, St. Petersburg 195251, Russia
}
\date{\today}

\begin{abstract}
We have developed a broadly-applicable approach that drastically increases the ability to accurately predict properties of complex atoms. We applied it to the case of Ir$^{17+}$, which is of particular interest for the development of novel atomic clocks with high sensitivity to the variation of the fine-structure constant and dark matter searches.
 The clock transitions are weak and very difficult to identity without accurate theoretical predictions. In the case of Ir$^{17+}$, even stronger electric-dipole ($E1$) transitions eluded observation despite years of effort raising the possibility that theory predictions are grossly wrong.  In this work, we provide accurate predictions of transition wavelengths and $E1$ transition rates in Ir$^{17+}$. Our results explain the lack of observation of the $E1$ transitions and provide a pathway towards detection of clock transitions. Computational advances demonstrated in this work are widely applicable to most elements in the periodic table and will allow to solve numerous problems in atomic physics, astrophysics, and plasma physics.
\end{abstract}

\maketitle


High resolution optical spectroscopy of highly charged ions (HCI) became a subject of much recent interest due to novel applications for the development of atomic clocks and search for new physics beyond the standard model of elementary particles \cite{Crespo2008,BDF10,BDFO11,KSCS18}. HCI optical clock proposals, fundamental physics applications, and experimental progress towards HCI high-precision spectroscopy were recently reviewed in \cite{KSCS18}. HCI have numerous
optical transitions between long-lived states suitable for development of clocks with very low uncertainties, estimated to reach $10^{-19}$ level \cite{BerDzuFla12,DerDzuFla12,DzuDerFla12a,DzuDerFla13E}. A particular attraction of HCI clock transitions is their  exceptionally high sensitivity to a  variation of the fine-structure constant $\alpha$ and, subsequently to dark matter searches \cite{BDF10,BDFO11,KSCS18}.

In many theories beyond the standard model, in particular those involving light scalar fields that naturally appear  in cosmological models, the fundamental constants become dynamical (i.e. varying) \cite{DP94,Uza11,Arv15,StaFla15,HMSS18,SafBudDem18}.
If  the fundamental constants, such as $\alpha$, exhibit space-time variation,  so are atomic spectra and  clock frequencies, which is potentially detectable with atomic clocks.  The dimensionless 
factor $K$ quantifies
the $\alpha$-variation sensitivity
\begin{equation}
\frac{\Delta E - \Delta E_{0}}{\Delta E_0} = K \frac{\alpha - \alpha_0}{\alpha_0},
\end{equation}
where $\alpha_0$ is the current value of $\alpha$ \cite{CODATA2014} and $\Delta E_0$ is the clock transition energy corresponding to $\alpha_0$. Experimentally,  variation of $\alpha$ is probed by monitoring the ratio of two clock frequencies with different values of $K$. Most of the currently operating atomic clocks have $|K|<1$, with the Yb$^+$ octupole transition having the highest $K=-6$ \cite{DzuFla09}. HCI transitions allow for much higher sensitivities, with $|K|>100$ making them particularly attractive candidates for these studies \cite{BDF10,BDFO11,KSCS18}.

It was  recently shown that coupling of ultralight scalar dark matter to the standard model leads to oscillations of fundamental constants and, therefore, may be observed in clock-comparison experiments \cite{Arv15,Til15,Hee16}. In addition, dark matter objects with large spatial extent, such as stable topological defects, would lead to
transient changes in fundamental constants that are potentially detectable with networks of clocks
\cite{DerPos14,StaFla15,Wci16,RobBleDai17,WciAblBel18}. These recent advances make development of novel clocks with high sensitivity to these effects particularly exciting. The sensitivity of optical clocks to $\alpha$-variation makes them also sensitive to  light scalar dark matter.

Recent development of quantum logic techniques for HCI
spectroscopy  in which a cooling ion, such as Be$^+$, provides sympathetic cooling as well as control
and readout of the internal state of the HCI ion \cite{SchVerSch15,Schmoeger2015RSI,LeoKinMic19}
made rapid progress in the development of  HCI clocks
possible. Recently, the spectra of Pr$^{9+}$ were  measured in an electron beam ion trap \cite{BekBorHar19}, and the proposed nHz-wide clock line was found to be at 452.334(1) nm.

One of the main remaining stumbling blocks towards development of many HCI clock proposals is the large uncertainties in theoretical predictions of the clock transitions, in particularly in cases with holes in the $4f$ shell (for example Ir$^{16+}$ and Ir$^{17+}$) or mid-filled $4f$ shell (Ho$^{14+}$).
While there are high-precision methods that allow to reliably predict HCI transitions in ions with 1-4 valence
electrons to a few percent or better \cite{SafDzuFla14}, approaches for the $4f$-hole systems are still in development stage, and theory accuracy is not well established.
While the $M1$ transitions in Ir$^{17+}$ between states of the same parity have been measured to good precision \cite{WCB15}, clock transitions or, in fact, any transition between opposite parity states were not yet identified.
These transitions were expected to be observed in recent experiments. Their predicted transitions rates \cite{Bekker} were well within the
experimental capabilities,
because $M1$ transitions  with much smaller transition rates have been observed.
Lack of such observations brought serious concerns about the accuracy of theoretical predictions.  In this work, we resolved this problem, for the first time definitively demonstrating ability to converge the configuration interaction (CI) in systems with a few holes in the $4f$ shell and place uncertainty bound on the results.
Our results explain the lack of observation of the $E1$ transitions, and provide a pathway towards detection of clock transitions.

We note that this work also serves as a basis for efficient treatment of systems with many valence electrons - present approach
 can be used for a large variety of applications beyond HCI calculations. Numerous problems in astrophysics and plasma physics require an accurate treatment of systems with many valence electrons, such as Fe. The lack of accurate theory predictions creates problems in applications involving almost all lanthanides and actinides as well as many other open-shell atoms and ions of the periodic table.
There are many other problems, besides HCI where our results will be useful, for example for neutral atom lattice clocks based on $4f^{14} 6s6p~ ^3\!P^o_0 - 4f^{13} 6s^2 5d ~(J=2)$  transition in Yb \cite{SafPorSan18,DzuFlaSch18}. Previous calculations could not reliably predict atomic properties of the $4f^{13} 6s^2 5d ~(J=2)$ state, such as the differential  polarizability and magic wavelengths of the clock transition.

\begin{figure}[t]
\centering
\includegraphics[width=80mm]{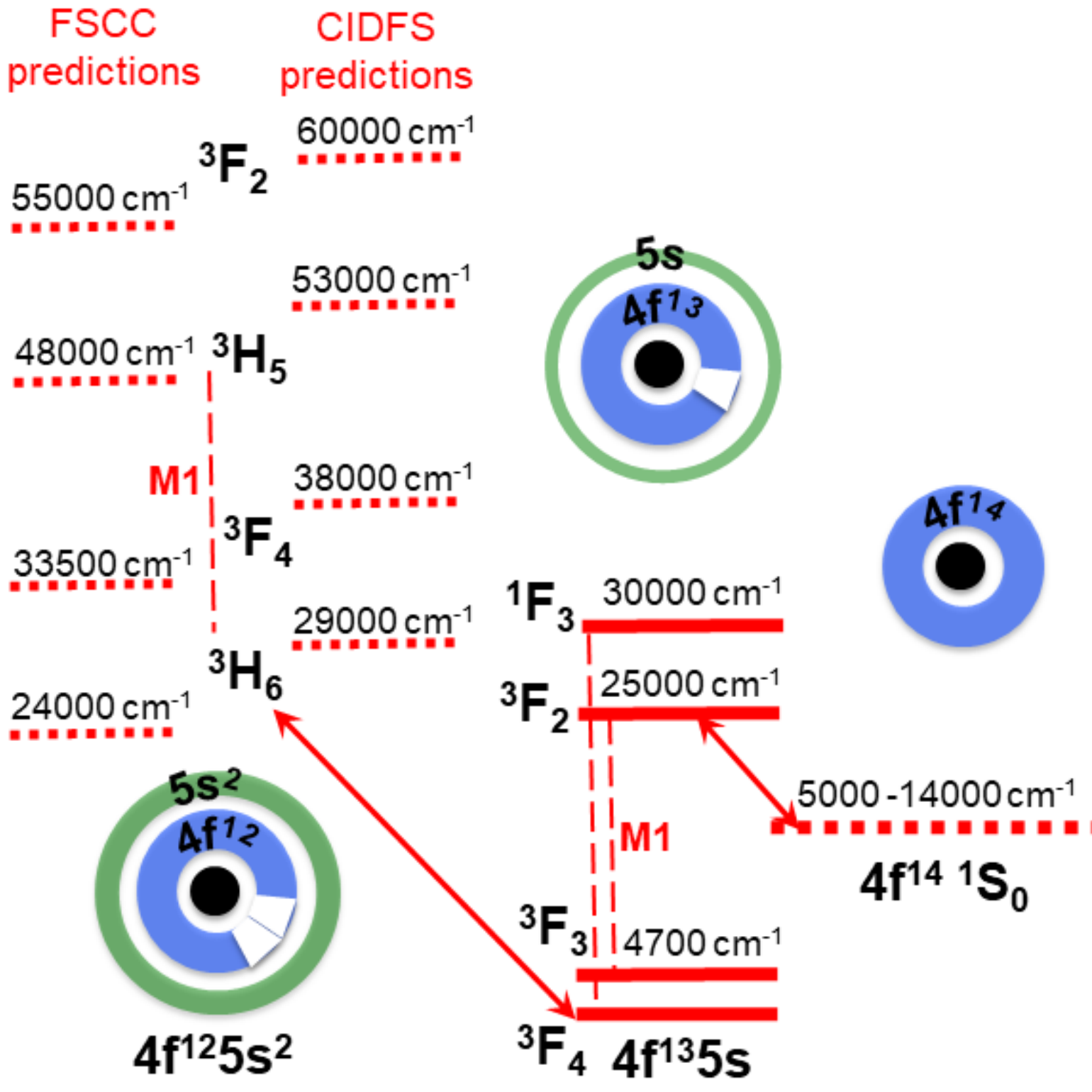}
\caption{Low-lying energy levels of Ir$^{17+}$ based on past CI Dirac-Fock-Sturm (CIDFS) and Fock space coupled cluster (FSCC) calculations~\cite{WCB15}. The scheme is not to scale. }
\label{fig1}
\end{figure}

The Ir$^{17+}$ ion has  [$1s^2$\dots $4d^{10}$] closed shells and a complicated energy level structure with $4f^{12}5s^2$, $4f^{13}5s$ and $4f^{14}$ low-lying levels shown in Fig.~\ref{fig1}. Prior calculations used the CI~\cite{BDFO11}, CI Dirac-Fock-Sturm (CIDFS)~\cite{WCB15}, Fock space coupled cluster (FSCC)~\cite{WCB15}, and the COWAN code~\cite{SFS15} methods. There is a reasonable agreement between them, from a few 100~cm$^{-1}$ to 1500~cm$^{-1}$, for the energy levels of the lowest  $4f^{13}5s$ states as all energies are counted from the ground state which has the same electronic configuration. However, there are very large, 5000 -- 13000~cm$^{-1}$ differences for all other levels. For convenience, we have shown the positions of the $4f^{12}5s^2$ levels predicted using CIDFS~\cite{WCB15} and FSCC~\cite{WCB15} calculations which are the most elaborate from all prior approaches. The CI results of \cite{BDFO11} place these levels much higher, by 5000 -- 7000~cm$^{-1}$.

\citet{BDF10} proposed to use the $4f^{12}5s^2\,^3\!H_{6} -\,4f^{13}5s\,^3\!F^o_{4}$ transition ($K=-22$) as a clock frequency reference. It is of $E3/M2$ transition type and can be enhanced by hyperfine-mixing with the $^3H_5$ state. They also note a possibility of using the $4f^{14}\,^1\!S_{0} - 4f^{13}5s\,\,^3\!F^o_{2}$ $M2$ transition which   may be induced by the hyperfine mixing with the $4f^{12}5s^2$~$^3\!P_{1}$ level. The particular attraction of this possibility is its very high (predicted to be $K=143$) sensitivity to the variation of $\alpha$. Fig.~\ref{fig1} illustrates the difficulty in predicting either one of these transition frequencies.

Nine of $M1$ transitions in Ir$^{17+}$ have been experimentally identified and measured at a ppm level \cite{WCB15}, including three $M1$ transitions shown in Fig.~\ref{fig1} by vertical dashed lines. The main puzzle is the lack of observation of two weak $E1$ transitions  \cite{Bekker} between the even and odd levels, i.e. $4f^{12}5s^2\,\,^3\!F_{4} - 4f^{13}5s\,\,^3\!F^o_{3,4}$ transitions. The theoretical determination of the odd level splittings is much more reliable than the odd-even energy difference, and observation of the $E1$ transitions would have allowed to determine the wavelength of the proposed  $4f^{12}5s^2\,\,^3\!H_{6} - 4f^{13}5s\,\,^3\!F^o_{4}$ clock transition with good precision. Moreover, it would provide a much needed test of theory models improving the prediction for the $4f^{14}\,\,^1\!S_{0}$ state.

We start from solutions of the  Dirac-Hartree-Fock equations in the central field approximation to construct one-particle orbitals. We find that the best initial approximation is achieved by solving Dirac-Hartree-Fock equations with  partially filled shells, namely $[1s^2 ... 4d^{10}] 4f^{13} 5s$. The hybrid approach  that incorporates core excitations into the CI by constructing an effective Hamiltonian with the coupled-cluster method \cite{CICC} cannot be used with such initial approximation. Therefore, the inner shells have to be treated using the CI method leading to an exponential increase in the number of required configurations. While the weights of most configurations are small, we find that the number of important configurations is still very large.

The increased size of the valence space imposes much higher computational demands. To resolve this problem, we developed a
message-passing interface (MPI) code that pre-estimates the weights of a very large number of configurations using  perturbation theory (CI-PT approach \cite{ImaKoz18}). We also developed codes to analyze the results and identify and sort the most important configurations.  Finally, we developed a fast MPI version of the CI code,  as the resulting set of important configurations was still extremely large.

The CI many-electron wave function is obtained as a linear
combination of all distinct states of a given angular momentum $J$ and parity:
$\Psi_{J} = \sum_{i} c_{i} \Phi_i\,.$
The energies and wave functions are
determined from the time-independent multiparticle Schr\"odinger equation
$H \Psi_n = E_n \Psi_n.$

In order to definitively ensure the reliability of the theoretical calculations, we consider all possible contributions that may affect the accuracy of the computations. We ensure the convergence in all numerical parameters: number and type of configurations included in the CI,  size of the orbital basic set used to construct CI configurations,  quantum electrodynamics (QED) and  Breit corrections. We find that by far the largest effect comes from the inclusion of the inner electron shells into the CI and we studied this effect in detail.

\begin{figure}[htb]
\centering
\includegraphics[width=60mm]{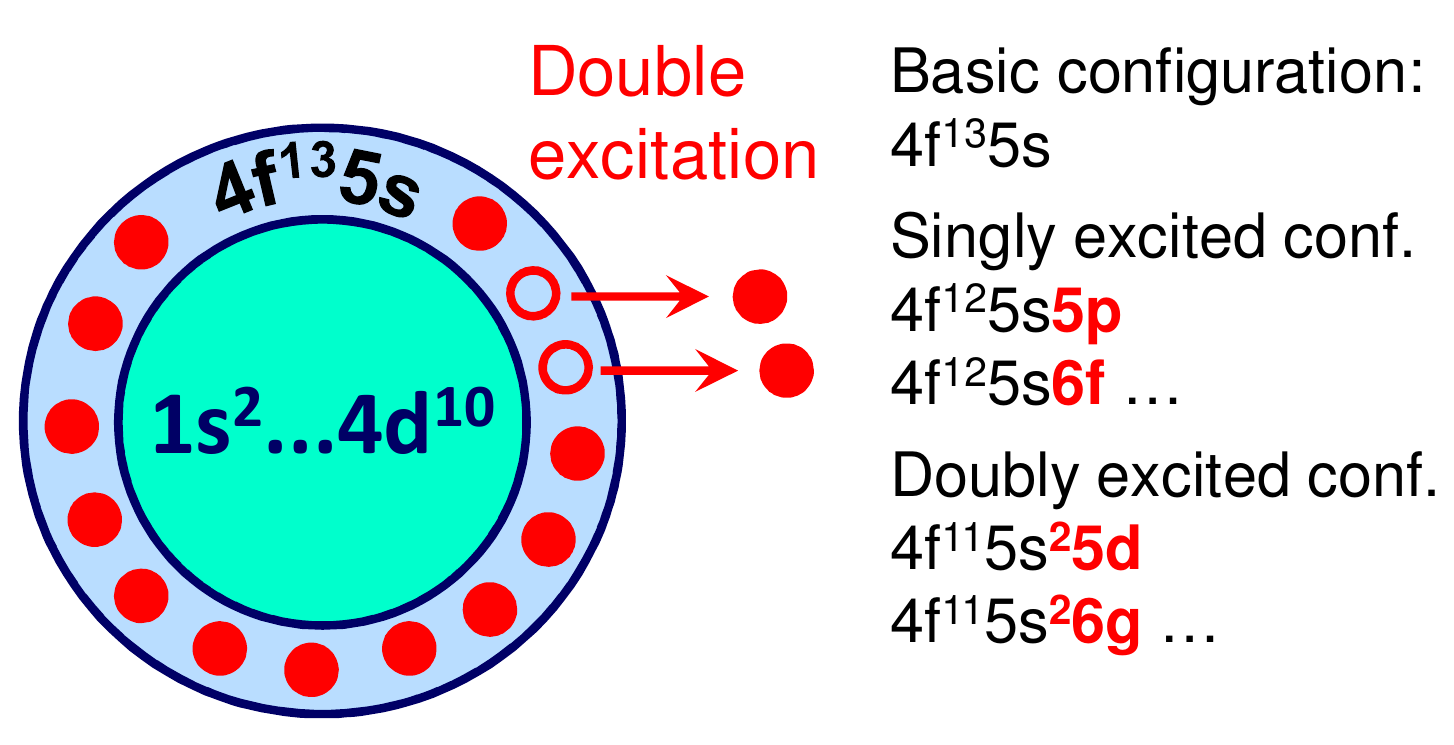}
\caption{Single and double excitations from the $4f^{13}5s$ configuration.}
\label{fig2}
\end{figure}

We start with the most straightforward CI computation that includes single and double excitations from the $4f$ and $5s$ valence shells, similar to \cite{BDFO11}.
 This is illustrated in  Fig.~\ref{fig2} which shows a few first configurations produced by exciting one and two electrons starting from the main $4f^{13}5s$ odd configuration. Excitations are allowed to each of the basis set orbitals. We begin with a basis set that includes all orbitals up to $7spdfg$ and discuss larger basis calculations below.

\begin{figure}[t]
\centering
\includegraphics[width=70mm]{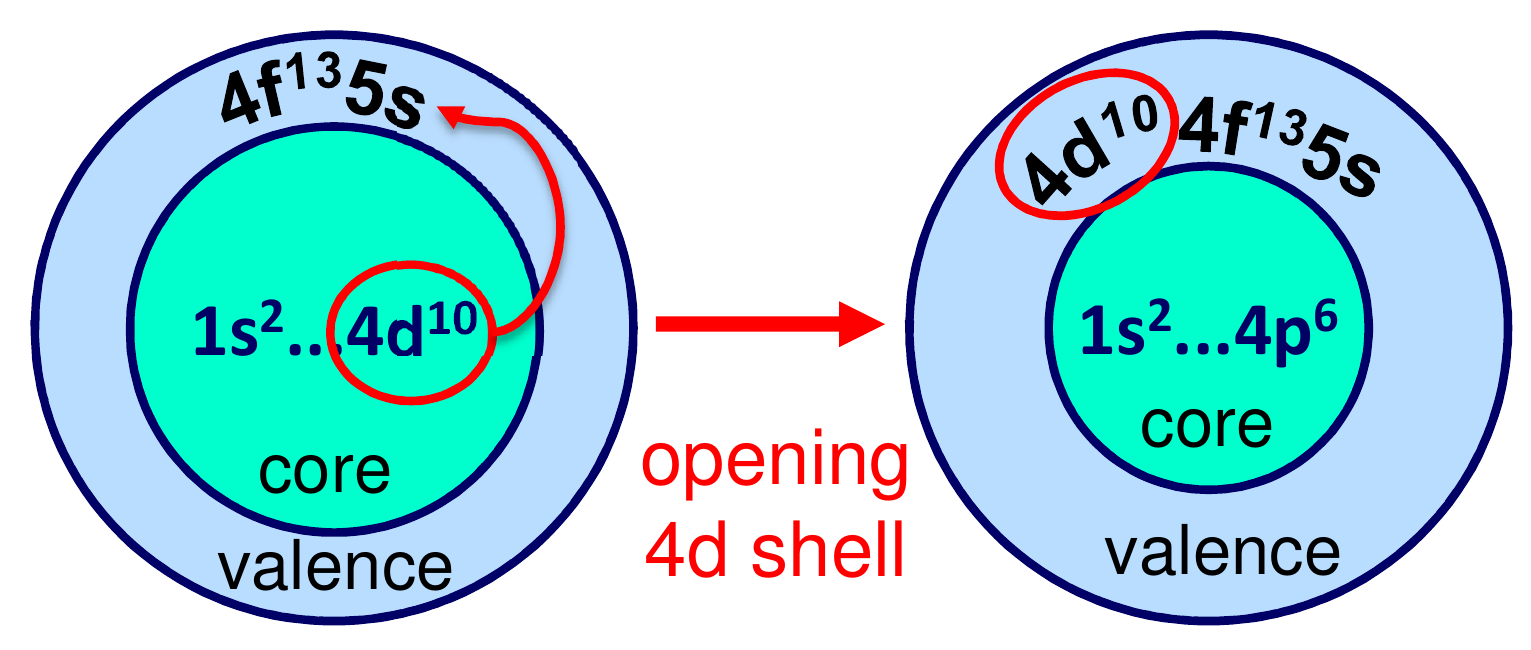}
\caption{``Opening'' the $4d^{10}$ shell - including it into the valence space.}
\label{fig3}
\end{figure}

Then, we ``open'' a $4d$ shell as is illustrated in Fig.~\ref{fig3}, i. e., include all $4d$ electrons into the valence space and allow excitations of any of the 24 electrons from the $4d^{10}4f^{13}5s$ shells to the same basis set orbitals up to $7spdfg$. We find drastic changes in the frequencies of all of the (odd-even) $E1$ transitions and the position of the $^1\!S_0$ level. This effect accounts for the difference between previous CI calculations \cite{BDFO11} which prohibited excitation of the $4d$ electrons and CIDFS calculations \cite{WCB15} which allowed it. In view of  such large contribution, we continued to include more and more electrons of the inner shells into the CI up to all 60 electrons.
Both single and double excitations  from the $4f, 4d, 4p, 4s$ and $3d$ shells are allowed and single excitations are included for all other shells. We tested that double excitation contributions are small for these inner shells and can be omitted at the present level of accuracy. The results, obtained with different number of shells included into the CI  are given in Table~\ref{tab1}. We note a very large contribution of the excitation from the $4s$ shell, which is the main sources of the difference between our results and CIDFS calculations \cite{WCB15}. All calculations in Table~\ref{tab1} are carried out with the same $7spdfg$ basis set.

Three different basis sets of increasing sizes, including all orbitals up to $7spdfg$, $8spdfg$, and $10spdfg$ were used to test the basis set convergence. Differences between results obtained with $7spdfg$ and $8spdfg$ basis sets do not exceed 264~cm$^{-1}$ for any level. All values are counted from the ground state. The difference between results obtained with $8spdfg$ and $10spdfg$ basis sets do not exceed 115~cm$^{-1}$ for any level.

We also
considered contributions of the triple excitations from the $5s4f$ shells and found them to be small at the present level of accuracy: $-600$~cm$^{-1}$ for the $^1\!S_0$ level and not exceeding $-351$~cm$^{-1}$ for all other levels. The sum of the corrections for a large $10spdfg$  basis, triples excitations, Breit correction beyond the Gaunt term, and QED corrections \cite{STY15,TKSSD16} is given in the column labelled ``Other'' in Table~\ref{tab1}. We note that these unrelated corrections substantially cancel each other. Based on the size of the inner shell contributions and all other corrections we estimate uncertainties of the final values for the even state to be on the order of 1000~cm$^{-1}$.
\begin{table*} [htb]
\caption{\label{tab1} \small Energies of Ir$^{17+}$ (in cm$^{-1}$) obtained using CI with  different  number of electronic excitations. Only configurations obtained by exciting $4f$ and $5s$ electrons are included
in the ``$5s4f$ only'' column. Excitation from the $4d$ shell are also included in the next column and the difference of the results is given in the column labelled ``$4d$ contr.''. Contributions of all other shells are given separately in the next columns. The results with all 60 electrons correlated by the CI are listed in the column ``All shells open''. Sum of all other corrections is given in the column labelled ``Other'' - see text for explanation.   }
 \begin{ruledtabular}
\begin{tabular}{lccccccccccccc} \multicolumn{2}{c}{Configuration}&
  \multicolumn{1}{c}{$5s4f$ only}&\multicolumn{1}{c}{$5s4f4d$}&\multicolumn{1}{c}{$4d$}& \multicolumn{1}{c}{$4p$}&
\multicolumn{1}{c}{$4s$}& \multicolumn{1}{c}{$3d$}&  \multicolumn{1}{c}{$1s2s3s$}&
\multicolumn{1}{c}{$3p$}&\multicolumn{1}{c}{$2p$}& \multicolumn{1}{c}{All shells} &\multicolumn{1}{c}{Other}&  \multicolumn{1}{c}{Final}\\
\multicolumn{4}{c}{}&
  \multicolumn{1}{c}{contr.} &\multicolumn{1}{c}{contr.} &\multicolumn{1}{c}{contr.} &\multicolumn{1}{c}{contr.} &\multicolumn{1}{c}{contr.} &\multicolumn{1}{c}{contr.} &\multicolumn{1}{c}{contr.} &\multicolumn{1}{c}{open}& \multicolumn{2}{c}{}\\
\hline \\[-0.7pc]
$4f^{13}5s$   & $^3\!F^o_4$ &  0     &     0     &      0 &     0   &      0  &     0     &   0    &    0   &0   &   0    & 0   &0 \\
              & $^3\!F^o_3$ & 4714   &   4745    &     31 &    15   &    14   &     8     &  -3    &    2   &0   & 4781   &-4   &4777 \\
              & $^3\!F^o_2$ & 25170  &   25095   &    -75 &    14   &    13   &    75     &  -2    &   25   &-4  & 25220  &-34  &25186 \\
              & $^1\!F^o_3$ & 30137  &   30253   &    116 &    51   &    33   &    73     &  -3    &   23   &-4  & 30426  &-31  &30395 \\ [0.5pc]
$4f^{14}$     & $^1\!S_0$   & 9073   &   14870   &   5797 &  -931   &  -1994  &   1097    & -240   &   -54  &9   &  12757 &-375 &12382 \\ [0.5pc]
$4f^{12} 5s^2$& $^3\!H_6$   & 36362  &   27813   &  -8549 &   460   &   1848  &   -403    &  183   &   294  &144 &  30339 &-56  &30283 \\
              & $^3\!F_4$   & 46303  &   37623   &  -8680 &    -5   &   1858  &   -410    &  184   &   251  &144 &  39645 &-81  &39564 \\
              & $^3\!H_5$   & 59883  &   51245   &  -8638 &   454   &   1858  &   -326    &  183   &   324  &143 &  53882 &-84  &53798 \\
              & $^3\!F_2$   & 68786  &   60036   &  -8751 &  -188   &   1690  &   -384    &  253   &   191  &64  &  61662 &-233 &61429  \\
              & $^1\!G_4$   & 69099  &   60056   &  -9043 &   165   &   1868  &   -322    &  184   &   304  &143 &  62397 &-136 &62261  \\
              & $^3\!F_3$   & 71963  &   63068   &  -8894 &   146   &   1836  &   -332    &  179   &   266  &146 &  65309 &-129 &65180  \\
              & $^3\!H_4$   & 91038  &   82254   &  -8784 &    78   &   1894  &   -245    &  187   &   340  &142 &  84650 &-126 &84524  \\
              & $^1\!D_2$   & 97473  &   87855   &  -9618 &  -110   &   1735  &   -334    &  270   &   177  &48  &  89639 &-366 &89273  \\
              & $^1\!J_6$   & 109332 &   99131   &  -10201&   268   &   1809  &   -304    &  171   &   212  &150 &  101437&-301 &101136  \\
\end{tabular}
\end{ruledtabular}
\end{table*}

\begin{table}[htb] 
	\caption{\label{tab2} Comparison of Ir$^{17+}$ $M1$ transition energies with experimental results \cite{WCB15}. Difference (in \%) of the other theory (FSCC and CIDFS) values from experiment \cite{WCB15} are given in the last two columns. }
	\begin{ruledtabular}
		\begin{tabular}{lcccccc}
			\multicolumn{2}{c}{Transition} &
			\multicolumn{1}{c}{Expt.} &
			\multicolumn{1}{c}{Present} &
			\multicolumn{1}{c}{Diff. \%} &
    		\multicolumn{1}{c}{FSCC} &
            \multicolumn{1}{c}{CIDFS}  \\
			\hline \\[-0.7pc]
			$4f^{13}5s$   & $^3\!F^o_2 -\, ^3\!F^o_3$ & 20711 &20409&	 1.5\%& 1.0\% & 2.6\%   \\
			              & $^1\!F^o_3 -\, ^3\!F^o_4$ & 30359 &30395&	-0.1\%& 0.5\% &-0.6\%  \\
			$4f^{12}5s^2$ & $^3\!H_5 -\, ^3\!H_6$     & 23640 &23515&	 0.5\%& 0.8\% & 1.4\%  \\
			              & $^3\!H_4 -\, ^1\!G_4$     & 22430 &22263&	 0.7\%& 0.5\% & 1.9\%   \\
			              & $^1\!G_4 -\, ^3\!F_4$     & 22949 &22697&	 1.1\%& 1.2\% & 1.3\%   \\
			              & $^1\!D_2 -\, ^3\!F_3$     & 23163 &24093&	-4.0\%&-2.0\% &-5.4\%    \\
			              & $^3\!F_3 -\, ^3\!F_4$     & 25515 &25616&	-0.4\%& 1.0\% &-0.1\%    \\
			              & $^1\!D_2 -\, ^3\!F_2$     & 27387 &27844&	-1.7\%& -0.1\%&-2.0\%  \\
			              & $^3\!H_4 -\, ^3\!H_5$     & 30798 &30726&	 0.2\%& -0.2\%& 1.7\%  			
		\end{tabular}
	\end{ruledtabular}
\end{table}

\begin{table}
\caption{\label{tab3} \small $E1$ $4f^{12}5s^2 - 4f^{13}5s$  radiative transition rates $A_{ab}$ of  Ir$^{17+}$ (in s$^{-1}$) obtained using CI with  different  number of electronic excitations: including excitations from the $4f5s$ electrons, then adding excitations from $4d$ and $4p$ shells. The final numbers include correlation of all 60 electrons. Final values of the transition wavelengths $\lambda$ (in nm) and reduced $E1$ matrix elements (in a.u.) are also listed. }
 \begin{ruledtabular}
\begin{tabular}{lcccccc} \multicolumn{1}{c}{Transition}&       \multicolumn{1}{c}{$\lambda$}& \multicolumn{1}{c}{$D$} & \multicolumn{4}{c}{Transition rate $A_{ab}$}\\
 \multicolumn{1}{c}{$a-b$}& \multicolumn{2}{c}{}&
  \multicolumn{1}{c}{$5s4f$}&\multicolumn{1}{c}{$+4d$}&\multicolumn{1}{c}{$+4p$}& \multicolumn{1}{c}{Final}\\
\hline \\[-0.7pc]
$^3P_2 -\,  ^3F^o_3$ &  91  &  4.1E-04 &  106  & 111  & 152 &  90   \\
$^3P_2 -\,  ^3F^o_2$ &  112 &  9.6E-04 &  727  & 458  & 276 &  269  \\
$^3P_2 -\, ^1F^o_3$ &  118 &  1.2E-03 &  1432 & 1101 & 254 &  333   \\
$^3H_4 -\,  ^3F^o_4$ &  118 &  1.6E-03 &  798  & 479  & 366 &  358   \\
$^1D_2 -\,  ^3F^o_3$ &  118 &  5.2E-04 &  9    &  4   &  91 &  65    \\
$^3H_4 -\,  ^3F^o_3$ &  125 &  1.8E-03 &  1325 & 891  & 347 &  369   \\
$^3F_3 -\,  ^3F^o_4$ &  153 &  1.3E-03 &  379  & 201  & 140 &  137  \\
$^1D_2 -\,  ^3F^o_2$ &  155 &  9.9E-04 &  515  & 277  & 103 &  104  \\
$^1G_4 -\,  ^3F^o_4$ &  160 &  1.8E-03 &  677  & 362  & 181 &  184   \\
$^3F_3 -\,  ^3F^o_3$ &  165 &  1.2E-03 &  579  & 319  & 85  &  90    \\
$^1D_2 -\,  ^1F^o_3$ &  169 &  1.2E-03 &  498  & 276  & 105 &  122   \\
$^1G_4 -\,  ^3F^o_3$ &  174 &  1.7E-03 &  376  & 209  & 123 &  129   \\
$^3F_2 -\,  ^3F^o_3$ &  176 &  1.5E-04 &  101  & 60   &  6  &  1.7   \\
$^3H_4 -\,  ^1F^o_3$ &  184 &  7.2E-05 &  216  & 0.3  & 0.2 &  0.18  \\
$^3F_4 -\,  ^3F^o_4$ &  252 &  4.9E-05 &  57   & 25   & 0.2 &  0.03   \\
$^3F_2 -\,  ^3F^o_2$ &  274 &  1.6E-04 &  60   & 26   & 0.3 &  0.47   \\
$^3F_3 -\,  ^1F^o_3$ &  287 &  4.3E-04 &  48   & 19   & 2   &  2.3    \\
$^3F_4 -\,  ^3F^o_3$ &  287 &  4.9E-04 &  64   & 30   & 1.2 &  2.2    \\
$^1G_4 -\,  ^1F^o_3$ &  313 &  1.8E-04 &  34   & 15   & 0.02&   0.2
\end{tabular}
   \end{ruledtabular}
\end{table}

$M1$ transition energies are compared with the experiment \cite{WCB15} in Table~\ref{tab2}; excellent agreement is observed with the exception of the $^1\!D_2 -\, ^3\!F_3$ transition. It is unclear if there might be an issue with the experimental identification, or the difference is due to the residual electronic correlations, as the contribution of the inner shells is particularly large here.

$E1$ transition rates of Ir$^{17+}$ (in s$^{-1}$) obtained using CI with different number of electronic excitations are given in Table~\ref{tab3}. While opening of the $4d$ shell drastically changed the energy levels, we found only small effect on the matrix elements; the differences in transition rates were caused by differences in energies. When the excitations from the $4p$ shells were included, we found only modest changes in the energies (see Table~\ref{tab1}), but a drastic reduction of the $E1$ matrix elements for a number of transitions. The multi-electron $E1$ transition rates are obtained from the one-body matrix elements, with the appropriate weights based on the mixing of the configurations. Allowing excitations from the $4p$ electrons accounted for previously omitted $4p-5s$ one-electron matrix elements,
whose role is particular important when the contributions from the one-electron $5s-5p$ and $4d-4f$
matrix elements are close in size but have the opposite sign and, respectively, essentially cancel each other. The final numbers include correlation of all 60 electrons, but the effect of all other shells for stronger transitions was relatively small.

Previous calculations of transition rate in Ir$^{17+}$ were only done with the FAC code \cite{Bekker} and did not include correlations besides the $4f5s$ electrons leading to incorrect prediction that  $^3\!F_4-\,  ^3\!F_4^o $ and $^3\!F_4 -\,^3\!F_3^o$ transitions should have been observable. In contrast, our calculations predict these transition rates to be very small, well outside of the detection range (see Table~\ref{tab3}). We identified a number of other transitions for the future $E1$ transition search,  where the transition rates are above 100~s$^{-1}$. We have calculated all of the E1 transitions between the states listed in Table~\ref{tab1}, also including $^3P_2$ states but only list the strongest transitions and a few representative examples where the transition rates   drastically change with the opening of the $4p$ shell.

 In summary, we have developed new MPI CI code that for the first time allowed us to correlate all 60 electrons in the framework of the CI approach. It can use arbitrary starting open-shell potential, reducing residual electron correlation effects. We were able to directly evaluate the contribution of each electronic  shell to the energies and matrix elements. Our calculations explained the failed search for the $E1$ transitions in Ir$^{17+}$: the transition rates of the two transitions that were searched for are well below detection threshold. We made reliable predictions of the $E1$ and clock transition wavelengths, with evaluation of their uncertainties and provided predictions of   $E1$ transitions sufficiently strong for the experimental detection. As illustrated by Table~\ref{tab1}, the energies of the $E1$ and clock $^3\!H_6 -\,^3\!F_4^o  $ transitions are strongly correlated, and as soon as any of the $E1$ transition wavelength is measured, we will be able to establish the clock transition energy with much higher precision.

 The method discussed here is very broadly applicable to
 many elements in the periodic table. Numerous problems in atomic physics, astrophysics, and plasma physics require accurate treatment of  open-shell systems similar to the one considered here. An exceptional speed up of the CI computations demonstrated in this work will allow to perform computations for other systems where reliable predictions do not yet exist. Our present computations were only limited by the computer memory resources presently available to us, and the largest run took less than 3 days on 80 CPUs. The work presented in this Letter, coupled with the development of new methods of efficiently selecting dominant configurations and larger computer resources, will eventually lead to
 accurate theoretical predictions for most elements of the periodic table.

\begin{acknowledgments}
We thank Jos\'e Crespo L{\'o}pez-Urrutia for helpful discussions and careful reading of the manuscript.
This work was supported in part by U.S. NSF Grant No.\ PHY-1620687 and Office of Naval Research, award number N00014-17-1-2252.
S.G.P., M.G.K., I.I.T., and A.I.B. acknowledge support by the Russian
Science Foundation under Grant No.~19-12-00157.
\end{acknowledgments}
%

\end{document}